# Second Harmonic Generation from a Single Plasmonic Nanorod Strongly Coupled to a WSe$_2$ Monolayer

Chentao Li†, Xin Lu†, Ajit Srivastava†, S. David Storm§, Rachel Gelfand§, Matthew Pelton§, Maxim Sukharev‡, ‖, Hayk Harutyunyan*†

†Department of Physics, Emory University, 400 Dowman Dr., Atlanta, Georgia 30324, United States

§Department of Physics, UMBC (University of Maryland, Baltimore County), 1000 Hilltop Circle, Baltimore, Maryland 21250, United States

‡College of Integrative Sciences and Arts, Arizona State University, Mesa, Arizona 85212, United States

‖Department of Physics, Arizona State University, Tempe, Arizona 85287, United States

Supporting Information Placeholder

**ABSTRACT:** Monolayer transition metal dichalcogenides, coupled to metal plasmonic nanocavities, have recently emerged as new platforms for strong light-matter interactions. These systems are expected to have nonlinear-optical properties that will enable them to be used as entangled photon sources, compact wave-mixing devices, and other elements for classical and quantum photonic technologies. Here we report the first experimental investigation of the nonlinear properties of these strongly-coupled systems, by observing second harmonic generation from a WSe$_2$ monolayer strongly coupled to a single gold nanorod. The pump-frequency dependence of the second-harmonic signal displays a pronounced splitting that can be explained by a coupled-oscillator model with second-order nonlinearities. Rigorous numerical simulations utilizing a non-perturbative nonlinear hydrodynamic model of conduction electrons support this interpretation and reproduce experimental results. Our study thus lays the groundwork for understanding the nonlinear properties of strongly-coupled nanoscale systems.

The development of cavity quantum electrodynamics (QED) has provided new control over light-matter interaction.[1] The Purcell effect was the first demonstration of such control, showing that spontaneous emission can be modified by changing the photonic density of states of the environment. When the rate, Ω, at which energy is exchanged between cavity photons and emitters (known as the vacuum Rabi frequency) is faster than any decay process in the system, the emitter excitation and cavity photons form new, hybridized states known as polaritons.[2-4] These polaritons have enabled novel phenomena such as control of the optical Stark effect[5], polariton lasing[6], polariton condensation,[7] and control of chemical reactivity.[8, 9] Besides these fundamental phenomena, the strongly coupled systems also have the potential to enable applications in optoelectronic devices, optical sensors, and quantum computing.[10, 11, 12-15]

To reach the strong coupling regime, early experiments typically used high-quality-factor dielectric cavities such as Fabry-Perot resonators[16, 17], photonic crystals[18], whispering-gallery-mode resonators[19], and distributed-Bragg-reflector cavities[20], with relatively large mode volumes restricted by the diffraction limit of light ~$(\lambda/2n)^3$. By contrast, metal plasmonic nanocavities such as single nanospheres[21] and nanorods (NR)[22] are not subject to the diffraction limit and provide deeply subwavelength interaction volumes.

Several material platforms have been used for strong coupling to plasmonic nanocavities, including quantum dots and molecular excitons.[17, 23-28] Recent progress in the fabrication and characterization of transition metal dichalcogenides (TMDs) has enabled several demonstrations of strong coupling between plasmons and excitons in these TMDs.[21, 29-35] A particular advantage of these materials is their large exciton binding energy, which enables strong plasmon-exciton coupling at room temperature.[36] Furthermore, the two-dimensional geometry of these materials results in large in-plane dipole moments in the interaction area, significantly increasing coupling with the cavity mode.

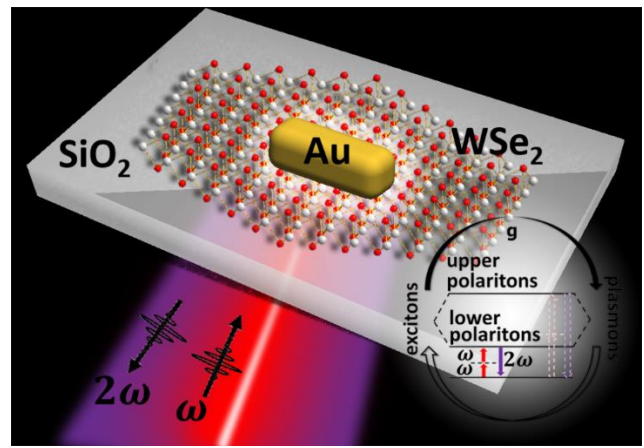

Figure 1. Schematic of the experimental setup. An isolated gold nanorod is strongly coupled to a monolayer of WSe$_2$ on a glass substrate. The second-harmonic signal at frequency 2ω, excited by a fundamental beam at frequency ω, is collected in the epi-illumination configuration.

So far, studies of these systems have been limited to their linear optical response. However, their nonlinear optical response has the potential to provide new routes for the development of nanoscale optoelectronic devices. For example, the nonlinearities may enable efficient entangled photon generation, compact wave mixing, and phenomena essential for optical quantum technologies and nanophotonic platforms. Additionally, the nonlinear signals can be potentially used to learn about the properties of underlying electronic states of the system, e.g. by probing the symmetries of polaritonic wavefunctions.

Here, we experimentally demonstrate second-harmonic generation (SHG) from single gold NRs coupled to a monolayer of WSe$_2$. The pump-frequency-dependent nonlinear signal shows a distinct spectral splitting. Numerical simulations and a simple analytic model show that this splitting can be attributed to formation of hybridized polaritonic states between plasmons and excitons in the TMD monolayer.

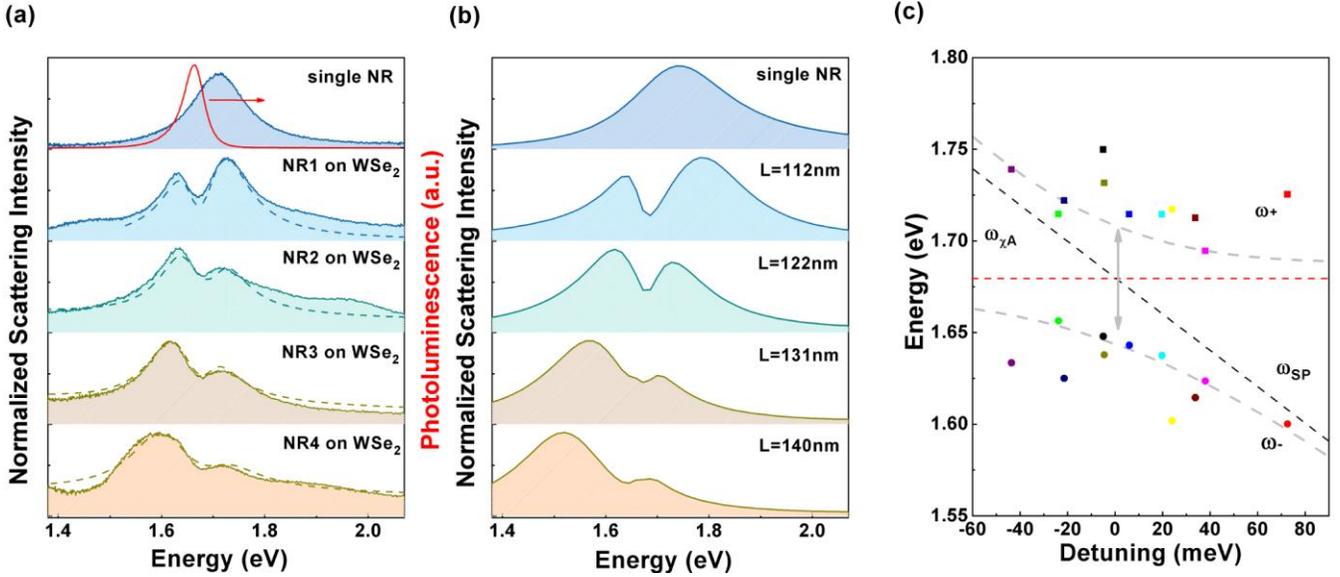

Figure 2. (a) Measured photoluminescence spectrum of a monolayer of WSe$_2$ (red, top panel), and measured dark-field-scattering spectra of an isolated gold nanorod (top panel) and of single nanorods coupled to the same WSe$_2$ monolayer (bottom panels). Dashed lines are fits using the linear coupled-oscillator model. (b) Calculated dark-field scattering spectra of an isolated gold nanorod (top panel, length is 112 nm and diameter is 40 nm) and of nanorods with different lengths coupled to a WSe$_2$ monolayer (bottom panels). (c) Frequencies of upper polaritons (solid squares) and lower polaritons (solid circles) extracted from fits to experimental data. Points with the same colors correspond to frequencies from the same scattering spectra. Grey lines are fitting results using the coupled-oscillator model, showing the anti-crossing behavior characteristic of strong coupling.

A schematic of the coupled system, which consists of a monolayer of WSe$_2$ and a single gold NR on top, is depicted in Figure 1. The sample is fabricated on a glass coverslip. (See Methods for details on sample fabrication).

The TMD monolayers are identified by raster scanning the sample and detecting the photoluminescence (PL) signal in a confocal configuration at room temperature (see Supporting Information Figure S1). The PL intensity is highly sensitive to the number of TMD layers, and only single monolayers a direct bandgap and strong emission;[37-39] an example of PL from a WSe$_2$ monolayer is shown in Figure 2 (a). A pronounced peak around 1.66eV is clearly observed, corresponding to emission from A excitons[37].

The gold NRs are designed to support longitudinal plasmonic resonances around 1.66eV, matching the A-exciton transition. Excitation of these plasmon modes give rise to confined in-plane electric fields at the surface of the WSe$_2$ flake, enabling coupling to excitons in the TMD. Figure 2 (a) shows examples of dark-field scattering spectra from an isolated gold NR and from individual NRs coupled to WSe$_2$ monolayers. The scattering spectra of the coupled systems show two peaks, as expected for the strong-coupling regime. (See Methods for details on optical measurements).

To unambiguously attribute the scattering spectral peaks to exciton-plasmon polaritons, Au NRs of different lengths, and thus different longitudinal plasmon frequencies, are investigated numerically (see Figure 2 (a) and Supporting Information Figure S2). Finite difference time domain (FDTD) simulations of the coupled systems' scattering spectra give results very similar to the experimental data, as illustrated in Figure 2 (b). (Parameters of the numerical simulations are provided in Supplemental Information.)

The linear scattering spectra can also be fit to a simple classical model, where the dipole moments, $\mu_{pl}$ and $\mu_{ex}$, of the plasmon and the exciton are represented as a pair of coupled harmonic oscillators:[40-42]

$$\ddot{\mu}_{pl} + \gamma_{pl}\dot{\mu}_{pl} + \omega_{pl}^2\mu_{pl} = F_o + g(\omega_{pl}d_{pl}/d_{ex})\mu_{em} \quad (1)$$

$$\ddot{\mu}_{ex} + \gamma_{ex}\dot{\mu}_{ex} + \omega_{ex}^2\mu_{ex} = g(\omega_{ex}d_{ex}/d_{pl})\mu_{pl} \quad (2)$$

where $\gamma_{pl}$ is the linewidth of the plasmon; $\omega_{pl}$ is the resonance frequency of the plasmon; $d_{pl}$ is the polarizability of the plasmon; $\gamma_{ex}, \omega_x$, and $d_{ex}$ are the corresponding terms for the exciton in WSe$_2$; and $g$ is the effective coupling strength between the plasmon and exciton. The external field is assumed to produce a driving force, $F_o$, on only the plasmon, because its polarizability is much greater than that of the exciton. The larger polarizability of the plasmon also means that only its dipole contributes to the scattering cross-section $\sigma_{scat}$:

$$\sigma_{scat} \propto \left|\omega^2 \mu_{pl}^{(1)}\right|^2 \quad (3)$$

with the steady-state solution for the dipole given by

$$\mu_{pl} = \frac{F_o(\omega_{ex}^2 - \omega^2 - i\omega\gamma_{ex})}{(\omega_{ex}^2 - \omega^2 - i\omega\gamma_{ex})(\omega_{pl}^2 - \omega^2 - i\omega\gamma_{pl}) - \omega_{ex}\omega_{pl}g^2} \quad (4)$$

Figure 2 (a) shows sample fits of the experimental data to Eqs. (3) and (4). In these fits, $\gamma_{pl}$ and $\omega_{pl}$ are contstrained to be within the range measured for scattering from gold NRs uncoupled to WSe$_2$, and $\gamma_{ex}$ and $\omega_{ex}$ are constrained to be near the values obtained from the photoluminescence spectrum of the bare WSe$_2$ layer; the only free parameters in the fits are thus the coupling strength $g$ and an arbitrary overall scaling factor.

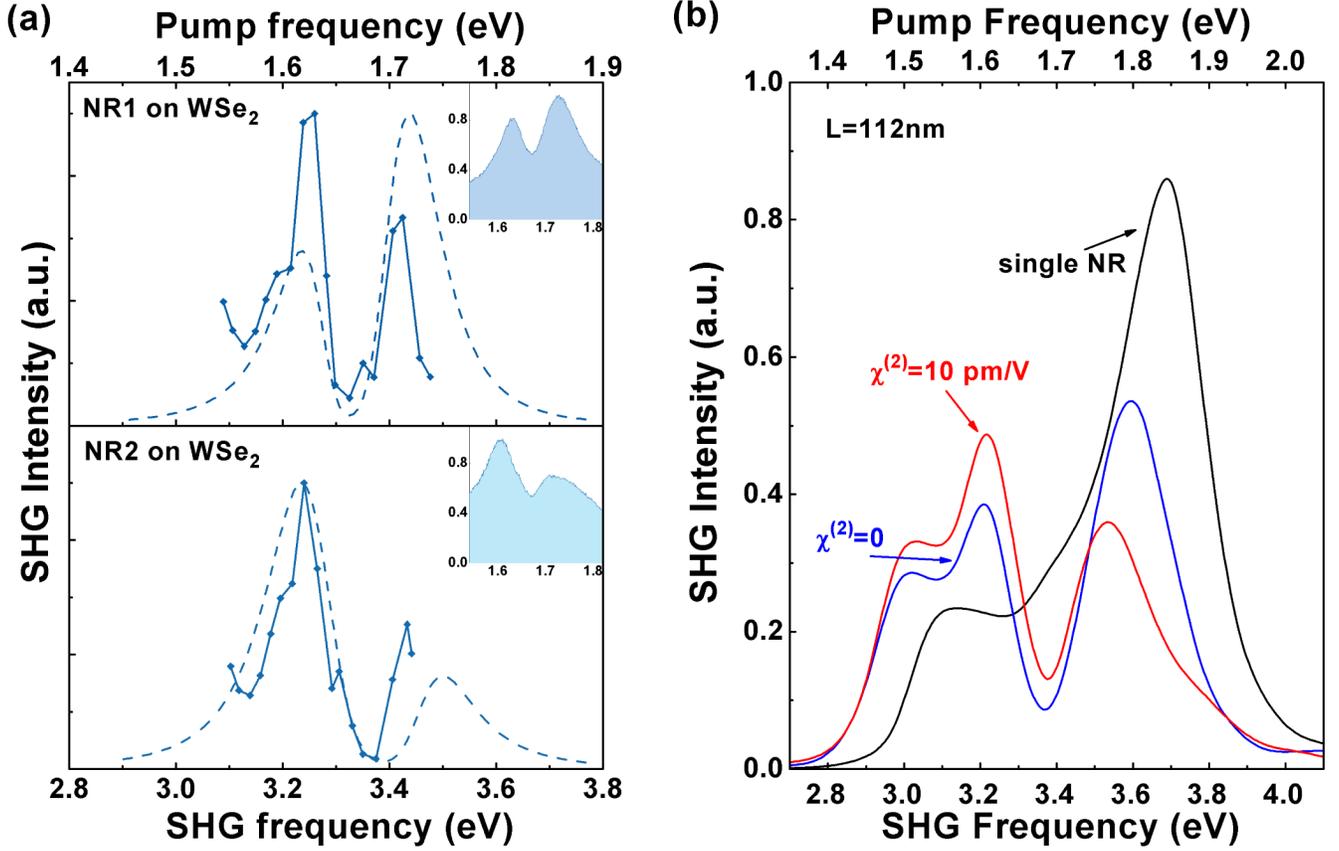

Figure 3. (a) Second-harmonic signals measured at different pump frequencies for single gold nanorods on a WSe$_2$ monolayer (solid lines). The corresponding linear scattering spectra are shown in the insets. Dashed lines show fitting results using the nonlinear coupled-oscillator model. (b) Calculated second-harmonic spectra for a single nanorod (length is 112nm, black) and the nanorod-WSe$_2$ coupled system using $\chi^{(2)} = 0$ (blue) and $\chi^{(2)} = 10$ pm/V (red) for WSe$_2$.

From the fitted parameters, the frequencies of the coupled plasmon-exciton modes can be calculated according to[42, 43]

$$\omega_\pm = \frac{1}{2}(\omega_{pl} + \omega_{ex}) \pm \sqrt{g^2 + \frac{1}{4}(\omega_{pl} - \omega_{ex})^2} \quad (5)$$

In Figure 2 (c), we plot $\omega_\pm$ as functions of the detuning $\delta \equiv \omega_{ex} - \omega_{pl}$, demonstrating the characteristic anti-crossing behavior. From this plot, we obtain an average coupling strength $g = 80 \pm 13$ meV. Comparing to the fitted ranges of plasmon linewidth, $\gamma_{pl} = 105 - 150$ meV and the exciton linewidth $\gamma_{ex} \approx 70$ meV, we can see that the coupled plasmon-WSe$_2$ system meets the strong-coupling criterion[4, 42]

$$g > \frac{1}{4}(\gamma_{pl} + \gamma_{ex}) \quad (6)$$

We now turn our attention to the nonlinear properties of the sample by performing SHG measurements in a confocal configuration. The spectral dependence of the integrated SHG signal as a function of excitation wavelength is shown for two representative systems in Figure 3 (a). These SHG spectra exhibit two distinct peaks, which match well with the positions of the peaks in the linear scattering spectra (Figure 3 (a), shaded areas). This indicates that the emitted second harmonic has its origins in the coupled system, rather than in either the gold NR or the WSe$_2$ separately.

The measured SHG spectra can be understood intuitively by extending the coupled-harmonic-oscillator model of Eqs. (1) and (2) to include second-order nonlinear terms:[44]

$$\ddot{\mu}_{pl} + \gamma_{pl}\dot{\mu}_{pl} + \omega_{pl}^2\mu_{pl} + a\mu_{pl}^2 = F_o + g(\omega_{pl}d_{pl}/d_{ex})\mu_{ex} \quad (7)$$

$$\ddot{\mu}_{ex} + \gamma_{ex}\dot{\mu}_{ex} + \omega_{ex}^2\mu_{ex} + b\mu_{ex}^2 = g(\omega_{ex}d_{ex}/d_{pl})\mu_{pl} \quad (8)$$

where $a$ and $b$ are proportional to the second-order nonlinear susceptibility of Au and WSe$_2$, respectively. These coupled nonlinear equations can be solved in the perturbation limit[45]

$$\mu_{pl} = \mu_{pl}^{(1)}e^{i\omega t} + \mu_{pl}^{(2)}e^{2i\omega t} + \cdots \quad (9)$$

$$\mu_{ex} = \mu_{ex}^{(1)}e^{i\omega t} + \mu_{ex}^{(2)}e^{2i\omega t} + \cdots \quad (10)$$

The linear terms $\mu_{pl}^{(1)}$ and $\mu_{ex}^{(1)}$ are the same as the solutions to the linear equations, Eqs. (1) and (2), whereas the second-order terms are given by

$$\mu_{pl}^{(2)} = \frac{aF_o^2(\omega_{ex}^2 - \omega^2 - i\omega\gamma_{ex})^2}{\left(\omega_{pl}^2 - (2\omega)^2 - i\omega\gamma_{pl}\right)\left[(\omega_{ex}^2 - \omega^2 - i\omega\gamma_{ex})(\omega_{pl}^2 - \omega^2 - i\omega\gamma_{pl}) - \omega_{ex}\omega_{pl}g^2\right]^2} \quad (11)$$

and

$$\mu_{em}^{(2)} = \frac{bF_o^2(g\omega_{ex}d_{ex}/d_{pl})^2}{(\omega_{ex}^2 - (2\omega)^2 - i\omega\gamma_{ex})\left[(\omega_{ex}^2 - \omega^2 - i\omega\gamma_{ex})(\omega_{pl}^2 - \omega^2 - i\omega\gamma_{pl}) - \omega_{ex}\omega_{pl}g^2\right]^2} \quad (12)$$

The second-order (hyper-Rayleigh) scattering cross-section is given by

$$\sigma_{SHG} \propto \left|\omega^2\mu_{pl}^{(2)}\right|^2 + \left|\omega^2\mu_{ex}^{(2)}\right|^2 \quad (13)$$

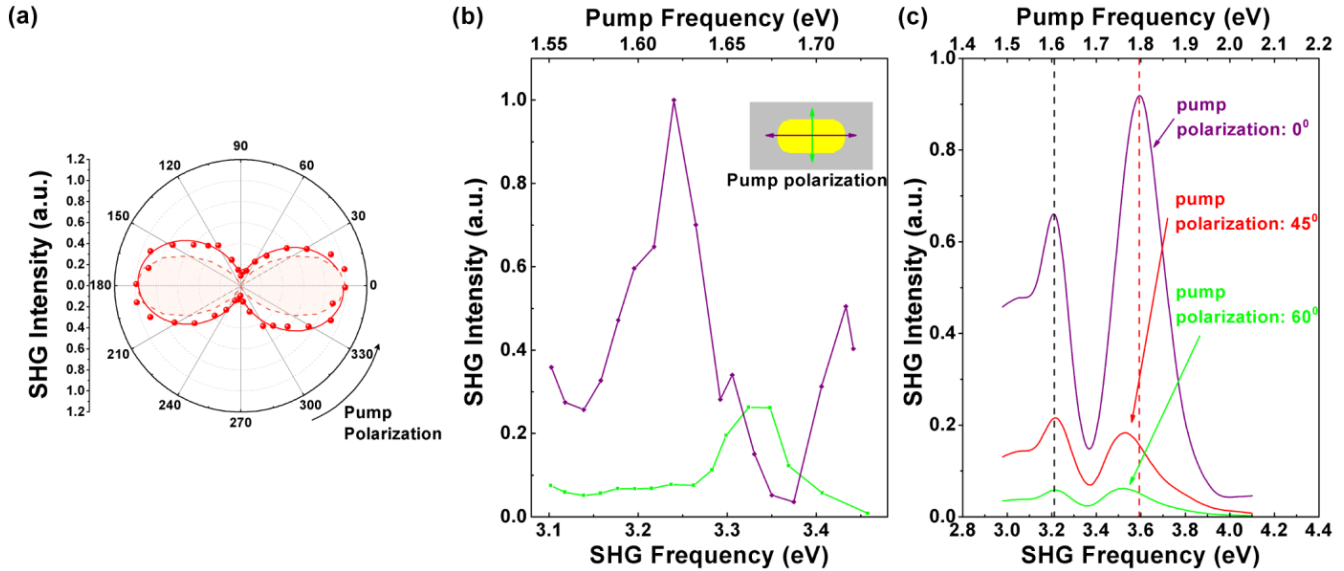

Figure 4. (a) Measured pump-polarization-dependent second-harmonic radiation pattern from the nanorod-WSe$_2$ coupled system (red dots) and a dipolar emission fit (solid red line). The numerical calculation results under the same condition are also shown (orange dashed line and filled area). (b) Measured second-harmonic spectra for the nanorod-WSe$_2$ system when pumping longitudinally (violet) and transversely (green) relative to the long axis of the nanorod. (c) Calculated second-harmonic spectra of a nanorod (length is 112nm) strongly coupled to WSe$_2$ when pumping with different polarization angles relative to the long axis of the nanorod.

For the same reason that we neglect linear scattering from WSe$_2$, we neglect the second term in Eq. (13) when comparing to experimental SHG spectra. This means that all the parameters in the fit to a given SHG spectrum are constrained by the fit to the corresponding linear spectrum, and the only free parameter is an arbitrary overall scaling factor. In practice, we account for possible calibration errors in the measurement of the SHG spectrum by including an offset and scaling factor for the frequency axis in the fit.

Fit results are shown in Figure 3 (a). The coupled-nonlinear-oscillator model shows good qualitative agreement with the measured SHG spectra. Quantitative differences are most likely due to errors in measuring SHG intensity at the edge of the spectral range of the detector used experimentally.

To confirm the validity of this simple physical picture, we also compare experimental SHG spectra to those obtained by numerically solving a non-perturbative fully vectoral hydrodynamic model coupled to Maxwell's equations[44]. Calculation results are shown in Figure 3 (c), and clearly reproduce the observed spectral splitting.

Including the second-order nonlinearity of WSe$_2$ in the simulations noticeably reduces the separation between the peaks in the SHG spectrum. This is in agreement with the coupled-nonlinear-oscillator model, which predicts a smaller splitting in the SHG spectrum due to the exciton dipole (Eq. (12)) than in the SHG spectrum due to the plasmon dipole (Eq. (11)). Since the total spectrum is a weighted sum of the exciton and plasmon spectra, increasing the fraction of SHG emitted by the exciton will reduce the overall peak separation.

All the calculated SHG spectra show an additional broad peak near 1.6 eV. Part of this peak can also be seen in the experimental spectra, near the edge of the experimentally accessible frequency range. This peak is attributed to local field enhancement at the tips of the rod. Further numerical details of the model, near-field distributions at the fundamental and harmonic frequencies, and additional simulations including third harmonic response are provided in the Supporting Information (Figures S5-S9).

Finally, we discuss the polarization dependence of the SHG spectra. The experimentally measured radiation pattern for a representative strongly-coupled system is shown in Figure 4 (a) along with results of the corresponding numerical calculation. A dipolar SHG emission pattern is observed, corresponding to the longitudinal plasmonic mode radiation.[46] In Figure 4 (b) we show SHG spectra recorded for longitudinal and transverse polarization of the pump beam. It can be clearly seen that a longitudinally-polarized pump excites the polaritonic states due to coupling between the longitudinal plasmon and the WSe$_2$ excitons. Meanwhile, the transverse pump produces signal only at the exciton peak, the intensity of which is 4 times smaller than that of the polariton peaks. Similar trends are seen for the numerical simulations as the pump polarization is rotated relative to the long axis of the NR, as seen in Figure 4 (c).

In conclusion, we report the experimental observation of Rabi splitting in the pump-frequency-dependent SHG signals from a strongly coupled system consisting of a gold nanorod and a monolayer of WSe$_2$. As described by FDTD simulations based on the non-perturbative hydrodynamic-Maxwell model, these polaritons reshape the SHG response of the system by creating local field enhancement at the polariton frequencies. The theory and simulations can be described using a simple, analytical model of two coupled nonlinear classical oscillators. Future work will be dedicated to extending the studies to nonlinear-optical effects beyond SHG, such as wave mixing and nonlinear extinction, and to plasmonic and excitonic systems with more complex symmetry properties. This, in turn, will pave the way for devices such as integrated entangled photon sources, room temperature quantum repeaters, and wave mixing elements.

**Methods**

Atomically thin layers of WSe$_2$ are mechanically exfoliated to a PDMS tape and then transferred to glass coverslips. Before transfer, the coverslips are ultrasonically cleaned in soap, de-ionized water, acetone, and IPA for 20 minutes in each solvent and are then dried using nitrogen. This method ensures the single crystalline structure of each WSe$_2$ flake.

To add gold NRs on top of the flake, we adopt the drop casting method described in Ref. 21. An aqueous solution of colloidal gold NRs with a diameter of 40 nm and an average length of 112 nm (Nanopartz Inc.) is 20-fold diluted with deionized water and sonicated to reduce aggregation. 1 μL of the diluted NR solution is then drop-cast onto the substrate with the $WSe_2$ flake and is washed off with deionized water after 1 min. These casting parameters result in around 30 NRs on each several-micron-sized $WSe_2$ flake. The NRs are functionalized as synthesized with a CTAB layer, which acts as a spacer between gold NRs and the $WSe_2$ flake to avoid charge transfer.

For PL measurements, a 50x objective (NA=0.5) is used to focus a 633-nm HeNe excitation laser beam on the sample and to collect PL signal in the epi-illumination configuration. PL maps are obtained by raster scan of the sample using an XY piezo stage and collecting the PL signal with an avalanche photodiode (APD). The scanning maps of several monolayer flakes are shown in the Supporting Information.

Scattering spectra are measured by focusing broadband light from a halogen lamp on the sample using a dark-field condenser lens. A 50x objective (NA=0.5) is used to collect the scattered light while omitting the transmitted light.

For SHG measurements, a Chameleon Ultra II Ti:sapphire femtosecond laser with a repetition rate of 80 MHz is used as the excitation source. A 60x oil immersion objective (NA= 1.4) is used to confocally focus the excitation beam and collect the SHG signals. To avoid the influence of higher-order plasmonic modes, a paraxial beam with a diameter of 2 mm at the objective-lens back focal plane is used, ensuring near-normal incidence of the excitation laser beam. On each NR candidate that shows a Rabi splitting in the dark field scattering spectra, nonlinear spectroscopy is performed by continuously tuning the pump from 700 nm to 800nm (3.1 eV – 3.6 eV) and recording the SHG spectrum for every excitation wavelength. To remove the broadband emission due to two-photon photoluminescence[47, 48], the raw SHG spectrum at each pump wavelength is fitted with a second order polynomial background and a sharp Gaussian peak (Supporting Information Figure S3). The area under the Gaussian peak is taken to be the intensity of the SHG signal. The quadratic dependence of the SHG signal on the excitation power is verified experimentally (Supporting Information Figure S4). To measure polarization-dependent SHG signals, we use a Berek compensator to rotate the linear polarization of the excitation beam while integrating the polarization of the emitted signal.

## ASSOCIATED CONTENT

### Supporting Information

Photoluminescence maps and spectra, dark-field scattering spectra, procedure for integrating experimental second-harmonic spectra, SHG power-dependence data, nonlinear hydrodynamic model details, discussion of calculated SHG spectra, third-harmonic generation calculations, spatial distribution of near fields at fundamental and SHG frequencies

## AUTHOR INFORMATION

### Corresponding Author

hayk.harutyunyan@emory.edu

### Author Contributions

H.H. conceived the project. C.L. performed the experiments. C.L., X.L. and A.S. fabricated the samples. D.S., R.G. and M.P. developed the analytical model. M.S. performed the numerical analysis. All authors contributed to the discussion of the results and the writing of the manuscript.

### Notes

The authors declare no competing financial interests.

## ACKNOWLEDGMENT

H.H. acknowledges support from the Department of Energy (DESC0020101). A.S. acknowledges support from NSF through the EFRI program-grant # EFMA-1741691 and NSF DMR award # 1905809. M.P. acknowledges support from the U.S. National Science Foundation (DMR-1905135). The computational part of this work is sponsored by the Air Force Office of Scientific Research under Grant No. FA9550-19-1-0009. The authors are grateful to Dr. Ruth Pachter and Prof. Stephan W. Koch for fruitful discussions pertaining to the optical response of $WSe_2$.

# Supporting information

## Second Harmonic Generation from a Single Plasmonic Nanorod Strongly Coupled to a WSe$_2$ Monolayer

Chentao Li†, Xin Lu†, Ajit Srivastava†, S. David Storm§, Rachel Gelfand§, Matthew Pelton§, Maxim Sukharev‡, ‖, Hayk Harutyunyan*†

†Department of Physics, Emory University, 400 Dowman Road, Atlanta, Georgia 30324, United States

§§Department of Physics, UMBC (University of Maryland, Baltimore County), 1000 Hilltop Circle, Baltimore, Maryland 21250, United States

‡College of Integrative Sciences and Arts, Arizona State University, Mesa, Arizona 85212, United States

‖Department of Physics, Arizona State University, Tempe, Arizona 85287, United States

*Corresponding author email: hayk.harutyunyan@emory.edu

# 1. Photoluminescence maps and spectra of exfoliated WSe$_2$ monolayers on glass substrates

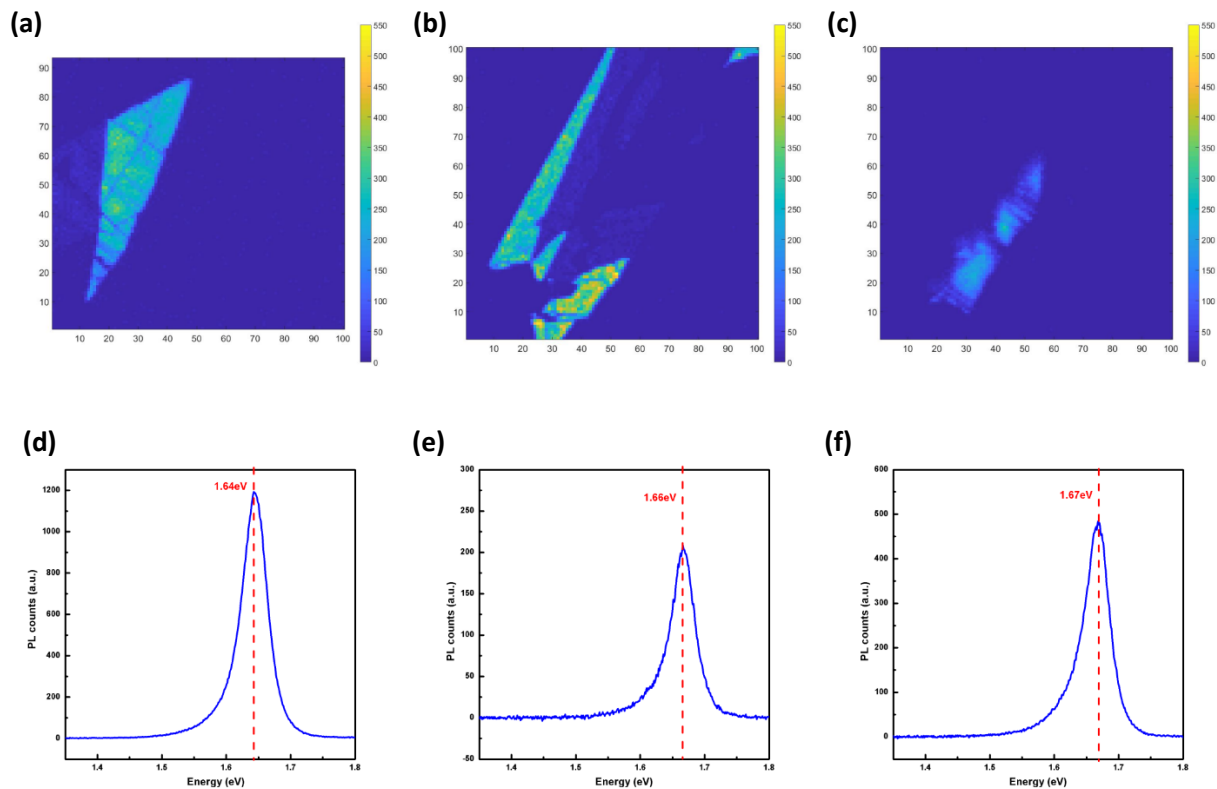

Figure S1. (a)-(c): Photoluminescence scanning maps of several WSe$_2$ monolayer. Each image shows an area of 100 $\mu$m × 100 $\mu$m. (d)-(e): Photoluminescence spectra of several WSe$_2$ monolayers. The peaks correspond to the A-exciton luminescence from WSe$_2$ and vary from 1.64 eV to 1.67 eV.

## 2. Dark-field scattering spectra for single nanorods coupled to WSe$_2$

Dark-field scattering spectra are measured on hundreds of candidate single nanorods coupled to WSe$_2$ to exclude spectra due to dimers, clusters, and dust. Besides the scattering spectra shown in Figure 2 (a), several additional spectra are shown in Figure S2.

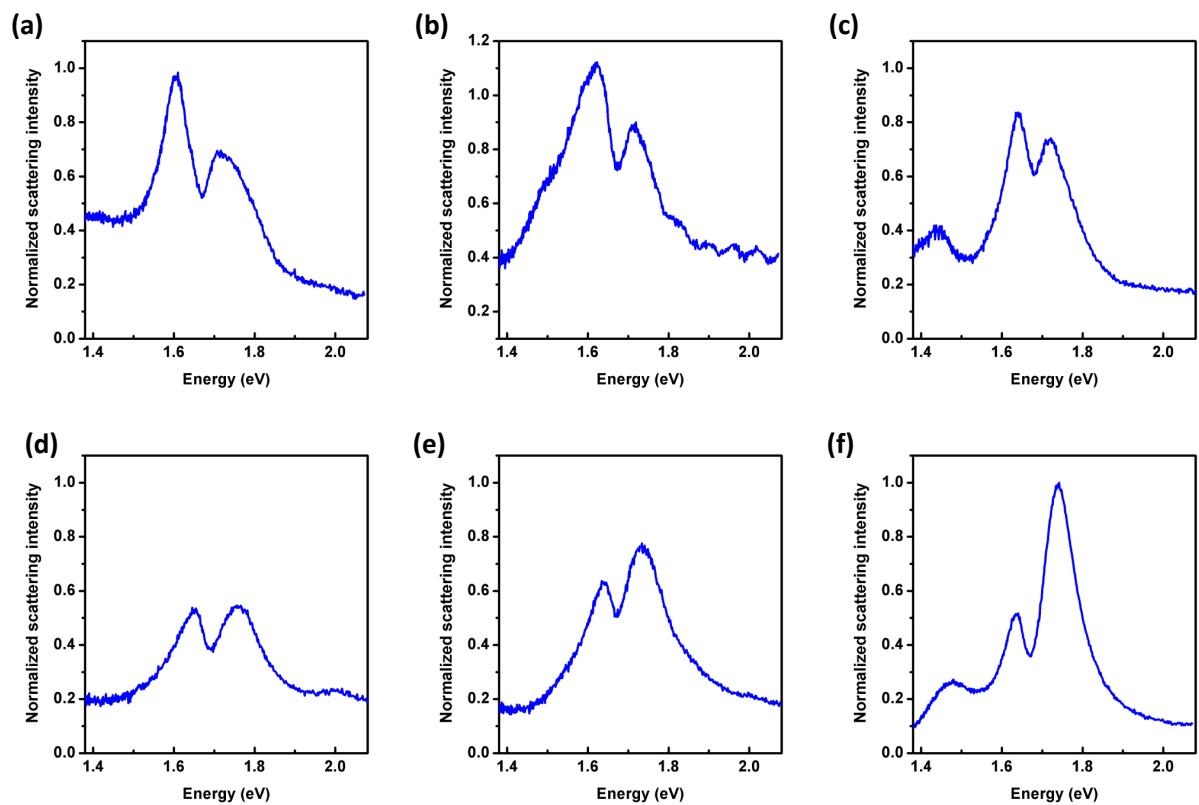

Figure S2. Dark-field scattering spectra for several nanorods coupled to a WSe$_2$ monolayer.

## 3. Procedure for integrating experimental second-harmonic spectra

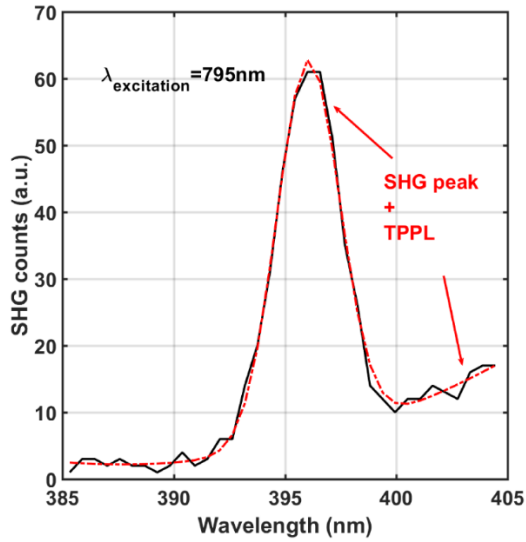

Figure S3. The black solid curve shows the raw second-harmonic signal from one nanorod-WSe$_2$ coupled system. The excitation wavelength is 795 nm and the time-averaged excitation power is around 250 $\mu$W. The red dashed curve is a fit to a Gaussian peak representing the second-harmonic signal plus a quadratic background representing two-photon photoluminescence. The area under the Gaussian peak is taken to be the intensity of the second harmonic signal.

## 4. Power dependence of second-harmonic signals

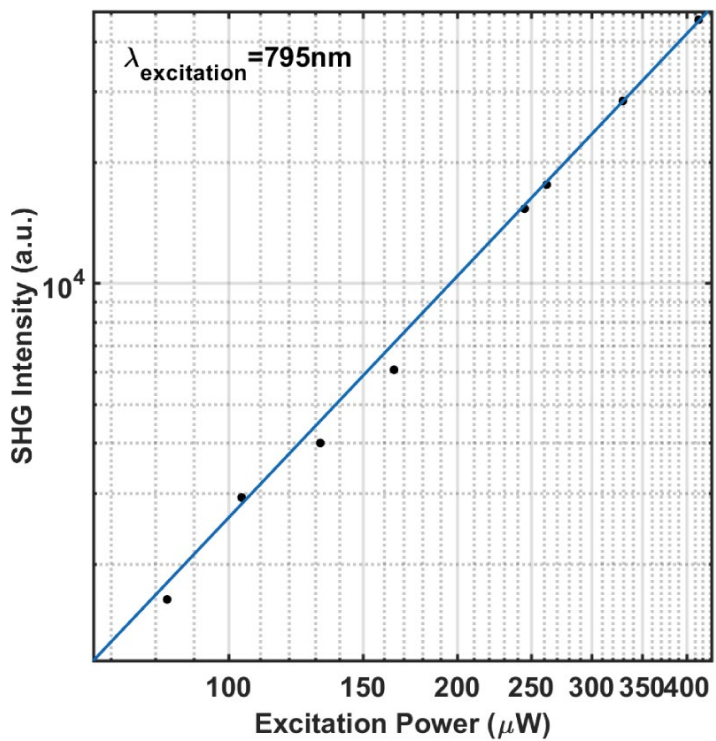

Figure S4. Excitation-power dependence of the second-harmonic intensity from gold nanorods coupled to a WSe$_2$ monolayer, and a power-law fit. The slope of the fit is $2.03 \pm 0.10$, in agreement with the expected slope of 2 for second-harmonic generation. The excitation wavelength is 795 nm. The excitation power is measured immediately after the objective without the sample.

## 5. Nonlinear hydrodynamic model of conduction electrons and its numerical implementation

The spatiotemporal dynamics of electromagnetic radiation is calculated using Maxwell's equations:

$$\varepsilon_0 \frac{\partial \vec{E}}{\partial t} = \frac{1}{\mu_0} \nabla \times \vec{B} - \frac{\partial \vec{P}}{\partial t},$$

$$\frac{\partial \vec{B}}{\partial t} = -\nabla \times \vec{E},$$

(S1)

where the macroscopic polarization, $\vec{P}$, is evaluated using the nonlinear hydrodynamic model for the conduction electrons. The latter relies on integration of equations for the electron velocity field, $\vec{u}$, and the electron number density, $n_e$:

$$\frac{\partial \vec{u}}{\partial t} + (\vec{u} \cdot \nabla)\vec{u} + \gamma_e \vec{u} = \frac{e}{m_e^*}(\vec{E} + \vec{u} \times \vec{B}),$$

$$\frac{\partial n_e}{\partial t} + \nabla \cdot (n_e \vec{u}) = 0.$$

(S2)

Here, $m_e^*$ is the effective conduction electron mass and $\gamma_e$ is a phenomenological decay constant. The current density, defined as

$$\vec{J} = \frac{\partial \vec{P}}{\partial t} = e n_e \vec{u},$$

(S1)

couples Eqs. (S1) and (S2). It is convenient to derive an equation for the polarization using (S2) and (S3):[1]

$$\frac{\partial^2 \vec{P}}{\partial t^2} + \gamma_e \frac{\partial \vec{P}}{\partial t} = \frac{e}{m_e^*}\left(n_0 e \vec{E} + \frac{\partial \vec{P}}{\partial t} \times \vec{B} - \vec{E}(\nabla \cdot \frac{\partial \vec{P}}{\partial t})\right) - \frac{1}{n_0 e}\left((\nabla \cdot \frac{\partial \vec{P}}{\partial t})\frac{\partial \vec{P}}{\partial t} + (\frac{\partial \vec{P}}{\partial t} \cdot \nabla)\frac{\partial \vec{P}}{\partial t}\right).$$

(S2)

The following parameters are used for gold: equilibrium number density $n_0 = 5.9 \times 10^{28}$ m$^{-3}$, plasma frequency $\omega_p = \sqrt{\frac{n_0 e^2}{\varepsilon_0 m_e^*}} = 7.039$ eV, and decay constant $\gamma_e = 0.181$ eV.

Equations (S1) and (S4) are numerically propagated in space and time using home-built finite-difference time-domain codes. The electric and magnetic field components are computed at different positions on the Yee lattice. The components of the macroscopic polarization are calculated at the same spatial positions as the corresponding components of the electric field. The entire computational domain is divided into a number of sub-domains, each carried by a single processor. We implement send/receive operations using message passing interface (MPI) subroutines on all six faces of each sub-domain. This parallelization methodology, known as a three-dimensional domain decomposition, renders the codes highly scalable.[10] Numerical convergence is achieved for a spatial resolution of 1 nm with a time step of 1.7 as.

It is important to note that, in time-domain simulations of harmonic generation processes, one needs to propagate the corresponding equations of motion for a considerable amount of time to achieve numerical convergence of the power spectrum. We found that power spectra converge after 500 fs if the system is driven by a 250-fs laser pulse with a peak amplitude of $2\times10^{-2}$ V/nm. The angular distributions of the second harmonic signal are highly dependent on the duration of the pump pulse, converging for pulses longer than 200 fs. Furthermore, to avoid mixing different nonlinear processes, the pump peak amplitude is chosen to ensure simulations are in the perturbative regime; i.e., the second-harmonic signal scales quadratically with the pump intensity. All simulations are performed on the AFRL/DSRC HPC clusters Mustang and Onyx using 1584 processors. Typical execution times of our codes vary between 20 (S1) – 30 (S2) minutes for linear simulations and 70 (S1) – 90 (S2) minutes to obtain second harmonic generation (SHG) results.

To incorporate the optical response of the $WSe_2$ monolayer into our numerical scheme, we adopt the Lorentz model with a frequency-independent second-order nonlinear susceptibility $\chi^{(2)}$. The value of $\chi^{(2)}$ is taken to be an adjustable parameter, with typical values varying between 1 pm/V[2] and 100 pm/V.[3] The numerical method that takes into account both linear dispersion and phenomenological nonlinear susceptibilities of different orders is adapted from Ref. [4].

The critical part in modeling of the experimental data is to ensure that the linear dispersion of $WSe_2$ is properly accounted for. It is well known that the environment influences the electronic structure of TMD materials rather significantly, which in turn may have a significant effect on their dielectric functions. When analyzing various dielectric functions for $WSe_2$ available in the literature, we used our experimental measurements as the main tool to identify the model most suitable for our work.

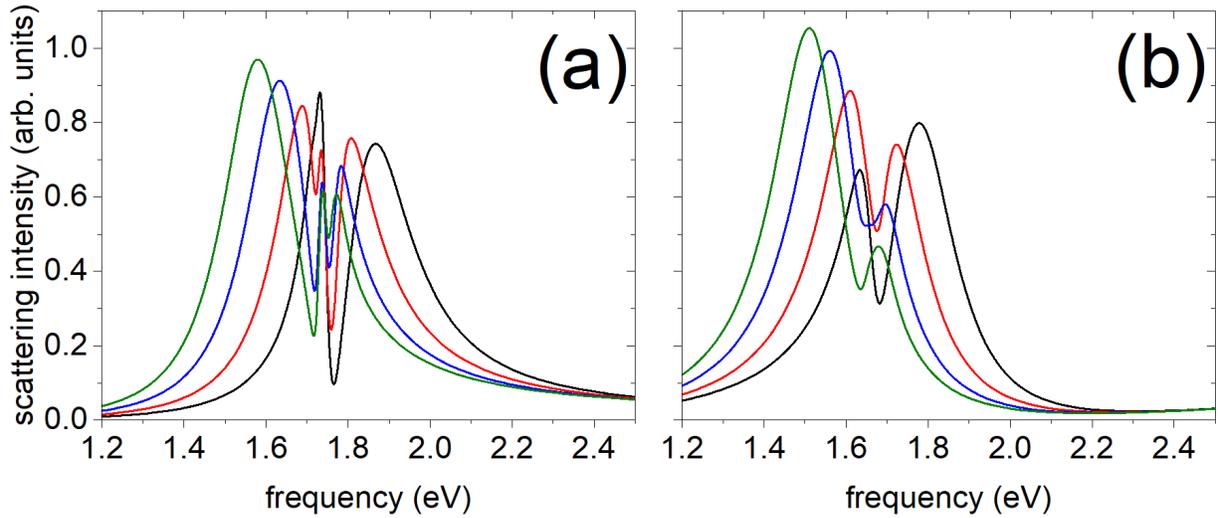

Scattering intensity as a function of the incident photon energy calculated for gold nanorods on top of a WSe$_2$ monolayer. The lengths of the rods are 112 nm (black), 122 nm (red), 131 nm (blue), and 140 nm (green). The diameter of all nanorods is 40 nm and the distance between the nanorods and WSe$_2$ is 5 nm. The system is excited by a pulse polarized along the nanorod's long axis. Panel (a) shows results using the dielectric function for WSe$_2$ from Ref. [5]. Panel (b) shows results approximating the dielectric function of WSe$_2$ as a single Lorentz oscillator with parameters from Ref. [6].

Fig. S5 compares linear spectra calculated using experimental parameters for WSe$_2$ from Ref. [5]. and from Ref. [6]. The scattering intensity obtained using the dielectric function of WSe$_2$ from Ref. [5] exhibits noticeable splitting reaching 133 meV for a 112-nm-long nanorod. Interestingly, the collective exciton mode is also observed for 122-nm and 131-nm nanorods as a third narrow spike at 1.74 eV. The physical nature of the third mode has been extensively discussed in the literature both theoretically and experimentally.[7-9] In our experiments, however, this feature has not been observed. By comparison, simulations based on the single Lorentz oscillator model with phenomenological parameters from Ref. [6] result in a Rabi splitting of 150 meV. Calculated widths of each polaritonic state closely match those seen in our experiments. Additionally, we performed fitting of the coupled oscillator model with experimental data and found that the closest match between simulations and experiments was obtained using parameters from Ref. [6]. We therefore use these parameters for all the calculations results reported in the main text.

# 7. More discussion of calculated second-harmonic signals

Fig. S6 shows normalized power spectra calculated for a single nanorod pumped on and off its longitudinal plasmon resonance.

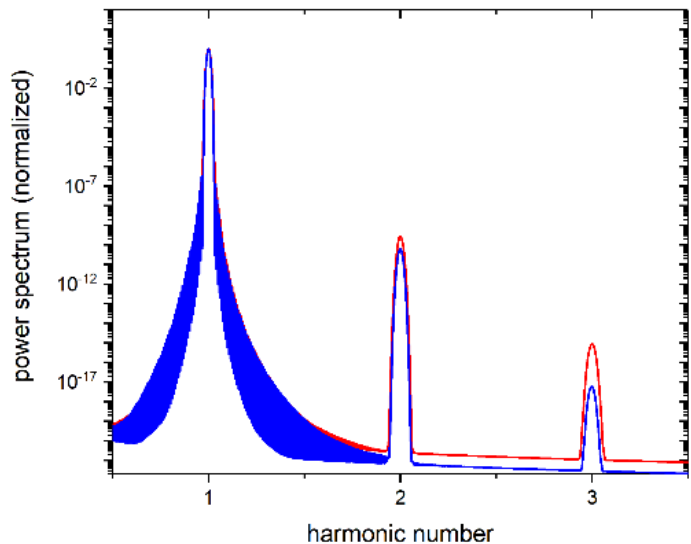

Figure S6. Normalized power spectrum as a function of harmonic number calculated for a single 112-nm-long gold nanorod pumped at its longitudinal plasmon resonance (1.83 eV, red) and off the plasmon resonance (2.00 eV, blue). The pump is polarized along the nanorod's long axis and is 250 fs long. Total propagation time of the calculations is 500 fs.

We note that, due to computational constraints, we limit the size of $WSe_2$ to a two-dimensional sheet with dimensions 500 nm by 500 nm. By comparison, in actual experiments, the dimensions of $WSe_2$ flakes are orders of magnitude greater than those of a single nanorod. Simulations carried out for slightly larger or smaller dimensions result in nearly identical data in the linear regime. However, nonlinear simulations are sensitive to the size of $WSe_2$.

To illustrate this, Fig. S7 shows the calculated SHG signal as a function of pump frequency for a stand-alone $WSe_2$ monolayer, for a stand-along gold NR, and for a $WSe_2$ monolayer coupled to a gold NR. The SHG spectrum from the $WSe_2$ monolayer alone has a peak near 1.7 eV, which results from direct-band-gap exciton. Additionally, the signal gradually increases with decreasing pump frequency. We ran several tests varying the size of $WSe_2$ to examine how it may affect SHG. The smaller the size of $WSe_2$ was, the steeper the low-frequency SHG response we obtained. The knee structure of the SHG signal for a single nanorod seen near 1.6 eV (dot-dashed line in Fig. S7) is attributed to the contribution of the local field enhancement at the tips of the rod. With the nanorod strongly coupled to a finite WSe$_2$ flake, the knee in the SHG response is pushed to lower pump frequencies.

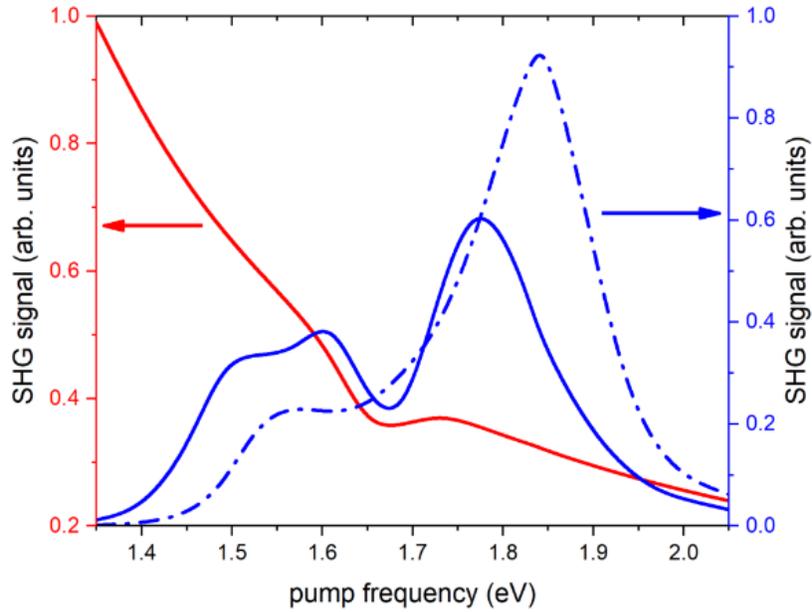

Figure S7. Second-harmonic signal as a function of the pump frequency. The red line (left vertical axis) shows the signal calculated for a stand-alone monolayer of WSe$_2$. The dot-dashed blue line (right vertical axis) shows signal for an isolated 112-nm-long nanorod. The solid blue line (right vertical axis) shows signal for the nanorod coupled to the WSe$_2$ monolayer. The nonlinear second-order susceptibility for WSe$_2$ is taken to be 10 pm/V.

## 8. Third-harmonic generation calculated for single nanorods and for nanorods coupled to WSe$_2$

As seen from Fig. S6, the pump also leads to significant third harmonic generation (THG). Although we did not study THG experimentally, it is worth exploring numerically. Fig. S8 shows THG results calculated for a stand-alone nanorod and for a nanorod coupled to linear WSe$_2$ (*i.e.*, we did not include the third order susceptibility of WSe$_2$ in our simulations). The results demonstrate that, even though WSe$_2$ is modeled as a linear Lorentz oscillator, THG is significantly altered due to strong coupling with the longitudinal plasmon of the nanorod. We note that third harmonics of the upper and lower polaritonic states are clearly visible. Interestingly, the maximum in the third-harmonic signal due to both polaritons is blue shifted with respect to the linear frequencies, with the blue shift of the upper polariton being greater than that of the lower polariton.

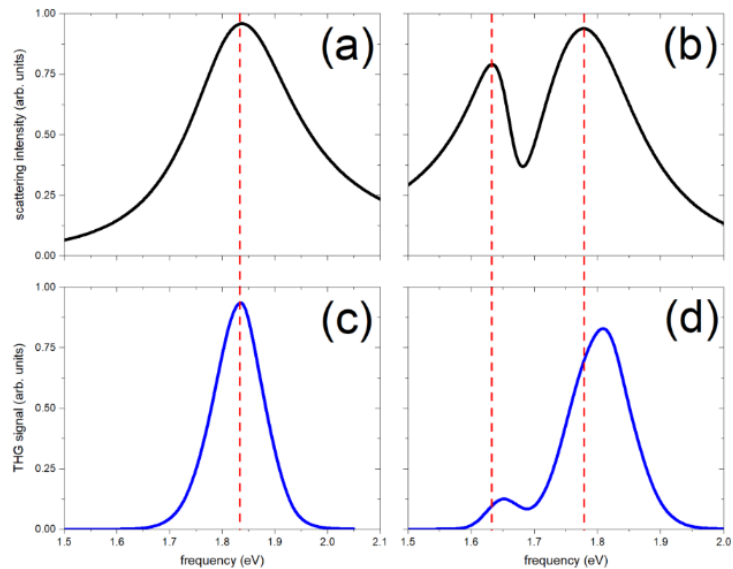

Figure S8. Panels (a) and (b) show linear scattering intensity as a function of frequency calculated for a single 112-nm-long nanorod (a) and for the nanorod on top of WSe$_2$ (b). Panels (c) and (d) show corresponding third-harmonic signals as functions of the pump frequency. Vertical red dashed lines indicate frequencies of the plasmon mode ((a) and (c)) and lower and upper polaritons ((b) and (d)).

## 9. Field maps for the strongly coupled nanorod-WSe₂ system

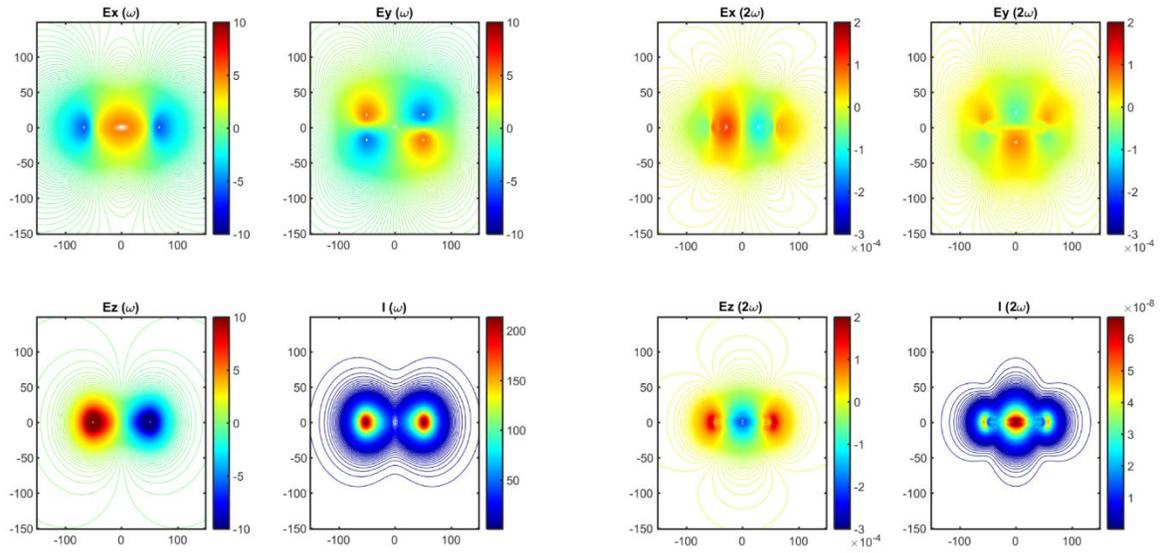

Figure S9. Local second-harmonic and fundamental field distributions and intensity distributions for the coupled nanorod-WSe₂ system. Headings on each panel indicate either a particular component of the electric field, $E_{x,y,z}$, or the intensity, $I$. The $x$ direction is along the long-axis of the nanorod, and the $z$ direction is normal to the substrate. The left four panels show the local field (in the units of enhancement) calculated at the longitudinal SPP frequency (1.83 eV), normalized by the magnitude of the incident field (or similarly for intensity). The right four panels show the electric field components evaluated at the second harmonic and normalized to the peak amplitude of the pump. All fields are calculated 10 nm above the surface of the nanorod. The nanorod's length is 112 nm and the incident field is polarized along the $x$ direction.